\documentclass[prb,aps,twocolumn,amssymb,amsmath,showpacs,,superscriptaddress]{revtex4}
\usepackage{graphicx} \newcommand{\figwidth}{.9\columnwidth}

\begin{document}
\title{Coulombian Disorder in Periodic Systems}

\author{Edmond Orignac}
\email{orignac@lpt.ens.fr} \affiliation{Laboratoire de physique th\'eorique
de l'\'Ecole Normale Sup{\'{e}}rieure, \\ CNRS UMR8549, 24, Rue Lhomond, 75231
Paris Cedex 05, France}

 \author{Alberto Rosso} \email{rosso@lptms.u-psud.fr}
\affiliation{ Laboratoire de physique th\'{e}orique et mod\`{e}les
statistiques CNRS UMR8626 \\ B\^{a}t. 100 Universit\'{e} Paris-Sud; 91405 Orsay Cedex,
France} 

\author{R. Chitra}
\email{chitra@lptl.jussieu.fr} \affiliation{Laboratoire de physique
th\'eorique de la mati\`ere condens\'ee CNRS UMR7600\\  Universit\'e de Pierre et Marie Curie,
4, Place Jussieu 75252
Paris Cedex 05 France}

\author{Thierry Giamarchi}
\email{giamarchi@physics.unige.ch} \affiliation{Universit\'e de
Gen\`eve, DPMC, 24 Quai Ernest Ansermet, CH-1211 Gen\`eve 4, Switzerland}

\begin{abstract}
  We study the effect of unscreened charged impurities on periodic
  systems.  We show that the long wavelength component of the
  disorder becomes  long ranged  and dominates static correlation
  functions. On the other hand, because of the statistical tilt
  symmetry, dynamical properties such as pinning remain unaffected. 
  As a concrete example, we focus on the
  effect of Coulombian disorder generated by charged impurities, on 3D
  charge density waves with non local elasticity.  We calculate the
  x-ray intensity and find that it is identical to the one produced by
  thermal fluctuations in a disorder-free smectic-A.  We discuss the
  consequences of these results for experiments. 
\end{abstract}
\maketitle

\section{Introduction} 
\label{sec:introduction}
The effect of quenched disorder on various condensed elastic systems is 
one of the fascinating problems in statistical mechanics. 
Examples of physical systems
range from domain walls in magnetic and ferroelectric
materials\cite{lemerle_domainwall_creep,tybell_ferroelectric}, contact lines of
of a liquid meniscus on a rough substrate\cite{moulinet_contact_line}, crack
propagation\cite{bouchaud_bouchaud}, to vortex lattices in type II
superconductors\cite{blatter_vortex_review,giamarchi_vortex_cargese}, charge density
waves (CDW)\cite{gruner_book_cdw,nattermann_brazovskii} and Wigner
crystals\cite{deville_wigner,williams_wigner,coupier_wigner,giamarchi_varenna}.
In these systems,  the competition between elastic interactions  which 
tend to impose some long range
order in the system and quenched disorder, leads to
the formation of glassy phases. Two broad classes of elastic systems can be
distinguished:  random manifold systems such as domain walls, contact
lines and cracks, and periodic systems such as  vortex lattices, charge
density waves, and Wigner crystals. The latter are characterized 
by a long range crystalline order in the absence of disorder and
thermal fluctuations. For these systems, a crucial question is whether
a weak disorder destroys entirely the crystalline order, or whether
some remnants of the underlying periodic structure remain observable.   
 
One of the earliest attempts to answer this question, was the
pioneering work by Larkin\cite{larkin_70} on vortex lattices. Using  a 
random-force model, he showed that due to the relevance of disorder 
in the renormalization
group sense, long range order was entirely destroyed below four
dimensions.  Above four dimensions, long range order persists as
disorder becomes irrelevant.  A similar conclusion was  reached 
by Sham and Patton for the
case of a CDW with short range elasticity
\cite{sham_peierls_disorder}, where, using an Imry-Ma
approach,\cite{imry_ma} they  concluded that long range order was impossible
in the presence of disorder below four dimensions.  The problem of
short range disorder in periodic systems with short ranged elasticity
was revisited in
Refs.~\onlinecite{villain_cosine_realrg,nattermann_pinning,giamarchi_vortex_short,giamarchi_vortex_long}.
It was argued that the periodicity present in systems like CDW and
vortex lattices plays a pivotal role in determining the physics of the
system in the presence of disorder.  More precisely, it was shown
that, though the disorder is relevant below four dimensions, due to 
the underlying periodicity  of the system  a quasi long-range order
persisted for dimensions  between two and four. This is in stark contrast to the
earlier results which predicted a total destruction of order.
The resulting phase, nicknamed Bragg glass phase, possesses both quasi
long range order and metastability and glassy properties.\cite{giamarchi_vortex_short,giamarchi_vortex_long} It was
further shown that the Bragg glass phase is stable to the formation of
defects.
\cite{giamarchi_vortex_long,gingras_dislocations_numerics,fisher_bragg_proof,zeng_fisher_nobglass2d}
Recent neutron scattering experiments on vortex lattices have
furnished clear evidence for the existence of such  a phase.\cite{klein_neutron}

A complication arises in charged  periodic systems due to the Coulomb repulsion,
which renders the elasticity
non-local.\cite{efetov_larkin_replicas,lee_coulomb_cdw,bergman_coulomb_disorder,chitra_wigner_hall}
This 
non local elasticity tends to rigidify the system, so that short range
correlated disorder
could  be irrelevant in dimension smaller than
four.\cite{chitra_vortex} For instance, within the
random force model, the correlation function of the displacement in
three dimensions displays a logarithmic growth indicating quasi-long
range order.\cite{efetov_larkin_replicas,bergman_coulomb_disorder,lee_coulomb_cdw} In fact,
when the periodic structure of the CDW is properly taken into account, 
the growth of the displacement correlation function is even weaker,
increasing only as $\log(\log(r))$ with the distance
$r$.\cite{rosso_cdw_long}      
 A second complication
arising from Coulomb interaction is that the disorder induced by
charged impurities has long-range correlations. This type of disorder
can exist in certain doped CDW materials\cite{rouziere_friedel_cdw}
such as $\mathrm{KMo_{1-x}V_xO_3}$. 

In this paper, we  study the effect of the competition of the
non-local elasticity produced by the Coulomb interaction with the
long-range random potential resulting from the presence of charged
impurities  on  the statics.   The paper is organized as follows: in
Sec.~\ref{sec:decomposition}, we introduce a decomposition of the
Coulomb potential on the Fourier modes of the periodic structure. With
this decomposition, we show that only the long-wavelength component of
the random potential i.e., forward scattering disorder, possesses long range correlations. Using
statistical tilt symmetry\cite{narayan_fisher_cdw}, we deduce that  due to the
short ranged nature of the  backward scattering terms engendered by the disorder, the
dynamical properties in the presence of charged
impurities are not qualitatively different from those in the presence of neutral short ranged impurities.
 In Sec.~\ref{sec:non-local-elasticity}, we consider the problem of
the 
CDW system, and we derive the non-local elastic Hamiltonian.  In
Sec.~\ref{sec:forward-scattering}, 
we derive the static displacement correlation functions and x-ray
intensity of the CDW with charged impurities and we highlight the
similitude of the latter to the x-ray intensity of smectic-A liquid
crystals subjected to thermal fluctuations.\cite{caille_smectic_xray} In
Sec.~\ref{sec:discuss}, 
we discuss the experimental significance of our result and suggest
that the smectic-like  correlations should be observable in experiments on
$\mathrm{KMo_{1-x}V_xO_3}$. Finally, we summarize the possible
behavior of the static correlators in a pinned Charge Density wave
according to the local or non-local character of elasticity and the 
presence or absence of charged impurities.

\section{Elasticity and Disorder in Periodic Systems}\label{sec:decomposition}

In this section, we discuss how Coulomb interactions affect
elasticity and disorder in periodic systems.  
For a  periodic elastic structure, the density  can be written as:
\begin{eqnarray}\label{density}
  \rho(\mathbf{r})=\rho_0(\mathbf{r}) + \sum_{{\bf G}} e^{i {\bf G}
  \cdot ({\bf r}-{\bf u}({\bf r}))},    
\end{eqnarray}
where $\rho_0({\bf r})=\rho_0(1-{\bf \nabla}\cdot{\bf u})$ describes
the density fluctuation arising from the long wavelength deformation
of the periodic structure and $\rho_0$ is the average density.  In the
second term, the vectors ${\bf G}$ belong to the reciprocal lattice of
the perfect periodic structure, and ${\bf u}({\bf r})$ represents a
slowly varying\footnote{by slowly varying, we mean that the Fourier
  components of $u$ are different from zero only for wavevectors much
  smaller than $|{\bf G}_{\text{min.}}|$.} elastic deformation of the
structure.\cite{giamarchi_vortex_long}  The quantities $e^{i {\bf G}
  \cdot ({\bf r}-{\bf u}({\bf r}))}$ describe fluctuations of the
density on the scale of a lattice spacing.  The low energy physics of
the periodic structure can be described in terms of a purely elastic
Hamiltonian which has the generic form for isotropic systems
\begin{equation} 
H_0= \int_{\bf r} \frac c2 (\bigtriangledown{\bf
    u})^2 
\end{equation} 
where $c$ is the elastic coefficient and $\int_{\bf r}$ is a shorthand
for $\int d^3{\bf r}$. This form can easily be generalized to
anisotropic systems.  Well known examples of charged periodic
structures are the Wigner
crystal,\cite{wigner,wigner_crystal,deville_wigner,williams_wigner,coupier_wigner}
charged colloidal crystals,\cite{colloids} and the charge density
waves.\cite{froehlich_cdw,peierls_inst,denoyer_cdw_ttf-tcnq}  In many
charged systems, unscreened Coulomb interactions are present:
\begin{eqnarray}
  H_{C} = \frac {e^2} 2 \int_{\mathbf{r},\mathbf{r'}} \frac{ \rho(\mathbf{r})
  \rho(\mathbf{r'})}{4\pi \epsilon |\mathbf{r} - \mathbf{r'}|},  
\end{eqnarray}
\noindent and  strongly affect  the elasticity  and dispersion of the
  compression modes of the system.  
Moreover, in the presence of charged impurities, the original charge
density on the lattice interacts with the charge impurity yielding:
  \begin{eqnarray} \label{eq:disorder}
   H_{\text{dis.}} =e^2 \int_{\mathbf{r},\mathbf{r'}} \frac{ \rho(\mathbf{r})
  \rho_{\text{imp.}}(\mathbf{r'})}{4\pi \epsilon |\mathbf{r} - \mathbf{r'}|},    \end{eqnarray} 
\noindent where $\rho_{\text{imp.}}$   denote  the impurity density.
Using the
  decomposition of the density~(\ref{density}), we now show that the
  Coulomb interactions  fundamentally modify
  only the long wavelength components of the elasticity and of the
  disorder energy.
  
 To better handle  the periodicity of the elastic structure, it is
 convenient to 
use  the  decomposition of the Coulomb interaction  in terms of the
reciprocal lattice vectors $G$. In three dimensions, this
 decomposition reads    
\begin{eqnarray}\label{eq:coulomb}
  \frac {1}{4\pi |{\bf r}|}&=&\int \frac{d^3 {\bf q}}{(2\pi)^3}
  \frac{e^{i{\bf q}\cdot {\bf r}}}{{\bf q}^2} \nonumber \\ 
&=& \sum_{{\bf G}} e^{i {\bf G}\cdot {\bf r}} V_{{\bf G}}({\bf r})
\end{eqnarray}
\noindent  where
\begin{eqnarray}
 V_{{\bf G}}({\bf r})= \int_{BZ} \frac{d^3 {\bf
  q}}{(2\pi)^3}  \frac{e^{i{\bf q}\cdot {\bf r}}}{({\bf q}+{\bf G})^2}, 
\end{eqnarray}
\noindent and $\int_{BZ}$ indicates that the integral is restricted to the
  first Brillouin zone. It is straightforward to check that $V_{-\bf G}({\bf
  r})=V_{\bf G}^*({\bf r})$.
    Using Eq.~(\ref{eq:coulomb}), the interaction term  $H_C$ can be
  rewritten as: 
  \begin{eqnarray} \label{eq:coulombenergy}
    H_C&=&\frac {e^2} {2\epsilon} \sum_{G\ne 0} \int_{\mathbf{r},\mathbf{r'}} V_{{\bf
    G}}({\bf r} -{\bf r'}) e^{i {\bf G}\cdot({\bf u}({\bf r}) -{\bf u}({\bf
    r'}))}\nonumber \\ 
&& +\frac {e^2} {2\epsilon} \int_{\mathbf{r},\mathbf{r'}} V_0({\bf r} -{\bf r'})
    \rho_0({\bf r})  \rho_0({\bf r'}).
  \end{eqnarray}
Note that due to the slow variation of  ${\bf u}({\bf r})$,  terms involving 
   the oscillatory factors $e^{i({\bf G}-{\bf G'})\cdot {\bf r}}$ can
   be dropped from the interaction.  
Let us first consider the term involving long wavelength fluctuations
   of the density.  
Since we are interested only in the long wavelength properties, we
  can replace the integration over the Brillouin zone in $V_0({\bf
  r})$.  by a Gaussian
  integration:
  \begin{eqnarray}
     \int_{BZ} \frac{d^3 {\bf q}}{(2\pi)^3} \to \int \frac{d^3 {\bf
     q}}{(2\pi)^3} e^{-a^2 {\bf q}^2},  
  \end{eqnarray}
\noindent with the parameter $a$ chosen so that $\pi/a \sim |{\bf
     G}_{\mbox{min.}}|$, ${\bf G}_{\mbox{min.}}$ being the reciprocal
     lattice vector having the shortest length. In this case, $V_0({\bf r})$
     can be obtained indirectly 
     by solving the Poisson equation with a Gaussian charge
     density and is found to be 
     \begin{eqnarray}
       V_{0}({\bf r})=\frac{1}{4\pi r}
       \mathrm{erf}\left(\frac{r}{2a}\right), 
     \end{eqnarray}
\noindent 
In the limit $r\gg a$, we recover the known result $V_0 \sim 1/(4\pi
r)$.  Clearly, the non-oscillating component of the Coulomb potential
remains long-ranged and tends to rigidify the system.
  
  It now remains to be seen whether the oscillating parts of the
  Coulomb interaction specified by $V_{{\bf G}}$ for ${\bf G}$ are
  long-ranged or not. We first note that the above trick of replacing
  the integration over the Brillouin zone by a Gaussian integral over
  the entire space is not applicable anymore, as it would introduce a
  spurious integration over a region where ${\bf G}+{\bf q}=0$. This
  would result in an (incorrect) $1/r$ behavior of $V_{\bf G \ne
    0}(r)$.  To obtain a correct estimate for $V_{\bf G}$ we replace
  the integral over the Brillouin zone by
\begin{eqnarray}
   \int_{BZ} \frac{d^3 {\bf q}}{(2\pi)^3} \to \int \frac{d^3 {\bf
     q}}{(2\pi)^3} F_{BZ}({\bf q},\epsilon),  
\end{eqnarray}
where $F_{BZ}(q,\epsilon)$ is an indefinitely derivable function with
a compact support contained in the first Brillouin zone (see
App.~\ref{app:decomp-coulomb} for an explicit form of $F_{BZ}(q,\epsilon)$).\footnote{A more straightforward
  approach would be to keep the hard cutoff at the edge of the
  Brillouin zone. Then, the function $V_{\bf G}$ would decay as
  $1/r^2$ with an oscillating prefactor. The same oscillation would be
  also obtained for a short ranged potential, and is only a
  consequence of the hard cutoff.} Obviously, for ${\bf G}\ne 0$,
$F_{BZ}(q,\epsilon)/|{\bf q}+{\bf G}|^2$ is also an indefinitely
differentiable function of compact support. A well-known
theorem\cite{gelfand64_theorem,reed75_mmmp} then shows that the
Fourier transform of $F_{BZ}(q,\epsilon)/|{\bf q}+{\bf G}|^2$ is
indefinitely differentiable and for $r\to \infty$ is $o(1/r^n)$ for
any $n>0$. This implies that the function $V_{\bf G}(r)$ is short
ranged.  Incorporating the above results in
Eq.(\ref{eq:coulombenergy}), we see that while the non oscillating
part of the Coulomb interaction modifies the long wavelength behavior
of the elasticity, rendering it non-local, the short ranged nature of
the oscillatory terms merely renormalizes the elastic coefficients. This is
explicitly  shown in App.~\ref{app:backscatt} for the particular case of a CDW.
The resulting non-local character of the elastic interactions modifies
strongly the static
and dynamic properties of the
system.\cite{chitra_wigner_hall,chitra_wigner_long,rosso_cdw_long,chitra_wigner_zero}

To understand the nature of the interaction with the charged impurities, we   
use the above procedure to   rewrite the random potential generated by the
impurities as:
\begin{eqnarray}
  U_{\mbox{imp.}}(r)&=& \int_{BZ} \frac{d^3 {\bf
  q}}{(2\pi)^3}  \frac{e^{i{\bf q}\cdot {\bf
  r}}\rho_{\text{imp.}}({\bf q})} {{\bf q}^2},  \nonumber \\ &=&
  \sum_{\bf G} e^{i {\bf G}\cdot {\bf r}} U_{{\bf G}}(r).  
\end{eqnarray}
Using this in Eq.(\ref{eq:disorder}), the interaction of the system with the random potential   is given by
\begin{eqnarray}\label{eq:disorder-general}
   H_{\text{dis.}}&=&\frac {e^2} {\epsilon} \sum_{G\ne 0} \int_{\mathbf{r},\mathbf{r'}} U_{{\bf
    G}}({\bf r})  e^{i
    {\bf G}\cdot{\bf u}({\bf r})} \nonumber \\  
&& +\frac {e^2} {\epsilon} \int_{\mathbf{r},\mathbf{r'}} U_0({\bf r})
    \rho_0({\bf r}), 
\end{eqnarray}
In Eq.~(\ref{eq:disorder-general}), the interaction of $\rho_0$ with
the random potential $U_0$ is called forward scattering, and the terms
containing $e^{i\mathbf{G} \cdot\mathbf{u(r)}}$ are called backward
scattering. This nomenclature originates in the theory of electrons in
1D random potential.\cite{giamarchi_book_1d} To calculate the disorder
correlation functions, we consider the case of Gaussian distributed
impurities where \cite{itzykson_stat2} $\overline{
  \rho_{\text{imp.}}({\bf G+q}) \rho_{\text{imp.}}({\bf
    G'+q'})}=(2\pi)^3 D\delta_{{\bf G},{\bf -G'}} \delta({\bf q}+{\bf
  q'})$, the parameter $D$ measuring the disorder strength.
Consequently, we find that for ${\bf G}\ne 0$:
  \begin{equation}
   \overline{U_{\bf G}({\bf r})U_{-\bf G}({\bf r'})} = D \int_{BZ} \frac{d^3 {\bf
  q}}{(2\pi)^3}  \frac{e^{i{\bf q}\cdot ({\bf r}-{\bf r'})}}{({\bf
  q}+{\bf G})^4},  
  \end{equation}
\noindent 
Using the same arguments as before, we infer that the correlations of
$U_{\bf G}({\bf r})$ are short ranged, as in the case of neutral
impurities, for all ${\bf G}$ except ${\bf G}=0$.  This implies that
the backward scattering terms induced by disorder are short-ranged and
the treatment of these terms within the replica or the
Martin-Siggia-Rose\cite{martin_siggia_rose,dominicis_dynamics} methods
is identical to the case of neutral or screened impurities.  However, the
${\bf G}=0$ component
  \begin{eqnarray}
    U_0({\bf r})=\int_{BZ}\frac{d^3{\bf q}}{(2\pi)^3} \frac{e^{i{\bf
    q}\cdot {\bf r}}}{{\bf q}^2} \rho_{\text{imp.}}({\bf q}), 
  \end{eqnarray}
  manifests power law decay of the forward scattering correlations.
  This term however can be gauged out by the statistical tilt
  symmetry\cite{narayan_fisher_cdw}, and affects mainly the static
  properties of the periodic system.  Typically, in periodic systems
  with both short range disorder and local  elasticity, the
  contribution of the forward scattering disorder can be neglected and
  it is the backward scattering that induces collective pinning and
  Bragg glass features like a quasi order in the static correlation
  functions.  Here, we have shown that even in the case of long range
  disorder, the backward scattering terms behave essentially like
  their short ranged (neutral impurities) counterparts. 
  However, the effect of the forward scattering terms on
  the correlation functions has to be studied carefully.  In the next
  section, we show that in the case of charged impurities in a charge
  density wave system, the forward scattering term strongly modifies
  the static correlation. Finally, we remark that our
  decomposition of the elastic energy and the  impurity potential is not
  exclusive to the Coulomb potential  and is 
  applicable to other long-range potentials.
   As a result, the conclusions of the present sections are expected to be
  valid for more general long range potentials. 
  
\section{Charge Density Waves}\label{sec:non-local-elasticity}
In this section, we re-derive the elastic Hamiltonian for a $d$
dimensional CDW with screened Coulomb interactions between the density
fluctuations at zero temperature.  We consider an incommensurate CDW,
in which the electron density is modulated by a modulation vector $Q$
incommensurate with the underlying crystal lattice. In this phase, the
electron density has the following form~\cite{nattermann_brazovskii}:
\begin{equation}
 \rho({\bf r}) = \rho_0 +\frac{\rho_0}{Q^2} {\bf Q}\cdot
 \nabla\phi({\bf r}) 
 +\rho_1 \cos( {\bf Q}\cdot {\bf r} +\phi({\bf r}))) \,,
\label{eq:rhoexp}
\end{equation}
where $\rho_0$, is the average electronic density (see
App.~\ref{app:cdw-density} for details). The second term in
Eq.~(\ref{eq:rhoexp}) is the long wavelength density  
and corresponds to variations of the density  over scales larger than $
Q^{-1}$ . The last oscillating term describes the sinusoidal deformation of the
density at a scale of the order of $Q^{-1}$ induced by the formation
of the CDW with amplitude $\rho_1$  and phase $\phi$. 
 
In the absence of Coulomb interactions, the low energy properties of
the CDW can be described by an effective Hamiltonian for phase
fluctuations.  For CDW aligned along the $x$ axis, i.e., ${\bf Q}=
Q\hat x$, this phase only Hamiltonian
reads\cite{fukuyama_cdw_pinning,gruner_revue_cdw,feinberg_cdw,maki_phase_hamiltonian}:
 \begin{eqnarray}
   \label{eq:hamiltonian_cdw_sr}
  H_0 = \frac{\hbar v_F n_c}{4\pi} \int
   d^3{\mathbf{r}}  \left[ 
(\partial_x \phi)^2  + %
\frac{v_y^2}{v_x^2} (\partial_y  \phi)^2 + 
\frac{v_z^2}{v_x^2} (\partial_z  \phi)^2   \right], \nonumber \\    
 \end{eqnarray}
 where $v_F$ is the Fermi velocity and $n_c$ is the number of chains
 per unit surface that crosses a plane orthogonal to $Q$.  The
 velocity of the phason excitations parallel to $Q$ is
 $v_x=(m_e/m_*)^{1/2} v_F$ with $m^*$ the effective mass of the CDW
 and $m_e$ the mass of an electron.  $v_y$ and $v_z$ denote the phason
 velocities in the transverse directions.  A crucial observation is
 that a deformation of $\phi(r)$ along $Q$ produces an imbalance of
 the electronic charge density which then augments the electrostatic
 energy due to Coulomb repulsion between density fluctuations.  We
 evaluate the contribution of Coulomb interactions screened beyond the
 characteristic length $\lambda$ which accounts for the presence of
 free carriers. This length diverges in the limit $T\to
 0$.\cite{wong_coulomb_cdw,virosztek_collective_cdw}  The
 electrostatic energy takes the form:
\begin{eqnarray}
\label{eq:electr}
H_C = \frac{e^2}{8\pi \epsilon}  \int d^d{\bf r}  d^d{\bf r'}  e^{-|{\bf r}- {\bf r}'|/\lambda}
\frac{\rho({\bf r}) \rho({\bf r}')}{|{\bf r}- {\bf r}'|} \,, 
 \end{eqnarray}
\noindent where we have assumed for simplicity an isotropic dielectric permittivity
$\epsilon$ of the host medium.\cite{efetov_larkin_replicas,lee_coulomb_cdw,kurihara_coulomb_phasons,wong_coulomb_cdw,brazovskii_longrange}
Due to the periodicity of the CDW system, we can use the decomposition
of the Coulomb potential derived in Sec.~\ref{sec:decomposition},
obtaining:
 \begin{eqnarray}
   \label{eq:electr-fourier}
   H_C &=& \frac{e^2 \rho_0^2}{8\pi\epsilon Q^2}  \int_{{\bf r},{\bf
       r'}}  \partial_x\phi({\bf r})  
\frac{e^{-|{\bf r}- {\bf r}'|/\lambda}}{| {\bf r}- {\bf r}'|} 
\partial_{x'}\phi ({\bf r}') \\
&& + \frac{e^2 \rho_1^2}{2\epsilon} \int_{{\bf r},{\bf r'}} \left[ V_{-Q}({\bf r}-{\bf r}') e^{i(\phi
  ({\bf r})-\phi ({\bf r}'))} + \text{c.c.}\right]\nonumber, 
 \end{eqnarray}
 In Eq.~(\ref{eq:electr-fourier}), we have neglected the contribution
 of the higher harmonics of the CDW.  Note that the oscillating
 terms, as discussed in App.~\ref{app:backscatt}, 
 only contribute to a renormalization of the
 coefficients in the short range elastic
 Hamiltonian~(\ref{eq:hamiltonian_cdw_sr}) and thus can be neglected.
 However, the contribution of the long-wavelength term has more
 dramatic effects and reads
\begin{eqnarray}
\label{eq:electr2-3D}
H_C &=& \frac{e^2 \rho_0^2}{ 2 \epsilon Q^2 } \int_{BZ}
\frac{q_x^2}{\lambda^{-2}+ q^2} |\phi(q)|^2,
\end{eqnarray}
in the three dimensional case. It is interesting to note that Coulomb
interactions generate a non local elasticity i.e., a $q$-dispersion in
the elastic constant.

The total Hamiltonian  in $d=3$ now reads,
\begin{eqnarray} 
\label{eq:anisotropyfinal} 
H_{\text{el.}}&=&H_0+H_C=\frac 1 2 \int G^{-1}(q) |\phi(q)|^2,  \\
 G^{-1}(q)&=& \frac{n_c \hbar v_F}{2\pi} \left[\frac{q_x^2}{({\bf q}^2
 +\lambda^{-2})\xi^2} + q_x^2 + \frac{v_y^2}{v_x^2} q_y^2 + \frac{v_z^2}{v_x^2}
 q_z^2\right]  \nonumber\
\end{eqnarray}
\noindent 
 where  the lengthscale $\xi$ is defined by:
 \begin{eqnarray}
   \label{eq:xi-definition}
   \xi^2 = \frac{n_c \hbar v_F}{2\pi e^2 \rho_0^2} Q^2\epsilon. 
 \end{eqnarray}
  Depending on the ratio $\lambda/\xi$, two regimes of behavior can be
 identified: (i) Short ranged elasticity: when $\lambda/\xi \ll 1$ the
 Coulomb correction to the short range elasticity is small even in the
 limit $q\to 0$ and hence can be neglected.\\ (ii) Long range
 elasticity: for $\lambda/\xi\gg 1$, the Coulombian correction to the
 short-range elasticity cannot be neglected. This regime is relevant
 at low temperatures, when the number of free carriers available to
 screen the Coulomb interaction is suppressed by the CDW
 gap.\cite{wong_coulomb_cdw,virosztek_collective_cdw} Mean field
 calculations show that this regime is obtained for temperatures
 $T<0.2 T_c$ where $T_c$ is the Peierls transition
 temperature.\cite{virosztek_collective_cdw}  In the following, we
 focus on regime (ii), and accordingly, we take $\lambda^{-1}=0$ in
 Eq.~(\ref{eq:anisotropyfinal}).

\section{Forward Scattering}\label{sec:forward-scattering}

As discussed in Sec.~\ref{sec:introduction}, the case of the short ranged
elasticity has been studied by various authors.  For charged periodic
systems with short range disorder and a non-local elasticity generated
by Coulomb interactions, it is known that $d=3$ becomes the upper
critical dimension for disorder and the displacement correlations grow
as ${ B}(r) = \log \log \Lambda
r$.\cite{chitra_vortex,rosso_cdw_long} 
Here, we study the effect of the long range disorder
on the static correlations of a charged periodic system.  Since, the
backward scattering terms generated by such a disorder are short
ranged, they lead to the same physics as that of short ranged disorder
with the corresponding nonlocal elasticity. These terms contribute a
$\log\log r$ term to the displacement correlations. However, in this
case a simple dimensional analysis shows that the forward scattering
terms generate the leading contribution to the correlation functions.
In the following, we calculate the contribution of the forward
scattering disorder to the displacement correlation function in the
$d=3$ CDW.

\subsection{Displacement correlation functions}
\label{sec:displacement}
The displacement correlation function is defined by: 
\begin{eqnarray}
B(r)&=&\overline{\langle(\phi(r)-\phi(0))^2\rangle} \,\nonumber \\
&=& \frac{2}{L^6} \sum_{\bf q}\ \overline{\left\langle \phi(q)
    \phi(-q)\right\rangle}  [1-\cos {\bf q}\cdot{\bf r}] \,. 
\end{eqnarray}
The calculation of the correlation induced by the forward scattering
disorder is analogous to the calculation of Larkin for the random
force model.\cite{larkin_70} 
Assuming an infinite screening length $\lambda$, the Hamiltonian reads:
\begin{eqnarray}
\label{eq:Hamiltonian2}
H&=& H_{\text{el.}}
+\frac{e^2}{4\pi \epsilon}  \int d^3{\bf r} d^3{\bf r'} 
     \frac{\rho_{\text{imp.}}({\bf r}) \rho({\bf r'})} {|r-r'|} \nonumber \\
\end{eqnarray}
Using Eq.~(\ref{eq:rhoexp}) in Eq.~(\ref{eq:Hamiltonian2}), we obtain
an expression of the form Eq.~(\ref{eq:disorder-general}). Keeping 
only the forward scattering term we get:
\begin{eqnarray}
\label{eq:Hamiltonian3}
H = \int \frac  {d^3q}{(2\pi)^3}\left[ \frac{G^{-1}(q)}{2} |\phi(q)|^2 +
  \frac{i\rho_0 e^2 q_x}{Q\epsilon q^2 } 
  \rho_{\text{imp.}}(-q) \phi(q) \right]. \nonumber \\     
\end{eqnarray}
Shifting the field $\phi$ 
\begin{eqnarray}
  \label{eq:gauge}
 \phi(q) = \tilde{\phi}(q)  + \frac{e^2
   \rho_0}{Q\epsilon} \frac{i
q_x G(q)}{q^2} \rho_{\text{imp.}}(q),
\end{eqnarray}
\noindent  brings the Hamiltonian (\ref{eq:Hamiltonian3}) back to the form of
   Eq.~(\ref{eq:anisotropyfinal}).  
The average over disorder now yields:
\begin{eqnarray}
\overline{\left\langle \phi(q) \phi(-q) \right\rangle} &=& \left\langle
\tilde{\phi}(q) \tilde{\phi}(-q) \right\rangle \nonumber \\
&&+
\frac{e^4 \rho_0^2}{Q^2\epsilon^2} \frac{q_x^2 G(q)^2}{q^4}
\overline{\rho_{\text{imp.}}(q) \rho_{\text{imp.}}(-q)}  
\nonumber\\
&=& L^3  \left[T G(q) + \frac{e^4 \rho_0^2}{Q^2\epsilon^2} \frac{q_x^2
  G(q)^2}{q^4} D\right] 
\end{eqnarray}
where $\langle \ldots \rangle$  and 
$\overline{\mbox{\ldots}}$ denote    thermal average and disorder
average respectively. 
Eq~(\ref{eq:gauge}) shows that even in the presence of
Coulombian disorder, the statistical tilt
symmetry\cite{narayan_fisher_depinning} is preserved.  
This implies  that  in the presence of backward scattering disorder, the
forward scattering term can be gauged out by (\ref{eq:gauge}),
and the contribution of the forward scattering disorder $B^{FS}$ is simply
added to the one obtained from the backward scattering
disorder, $B^{BS}$.\cite{rosso_cdw_short,rosso_cdw_long} 
We conclude
\begin{equation}
B^{FS}(r)= 2 D   \frac{e^4 \rho_0^2}{Q^2\epsilon^2}  \int  \frac{d^3q}{(2\pi)^3}  \frac{q_x^2
  G(q)^2}{q^4}  [1-\cos ({\bf q}\cdot{\bf r})] \,.
\end{equation}
We want to evaluate this integral for the case of $v_y=v_z=v_{\perp}$.
In the following we will use $q_{\perp}^2=q_y^2 + q_z^2$.  To obtain
the asymptotic behavior of $B(r)$ for $r\to \infty$ we need to
consider the $q \to 0$ limit of the integrand. The form of $G(q)$
suggests a scaling $q_x \sim q_{\perp}^2$ which then  allows us to consider
the integral:
\begin{eqnarray}
  F({\bf r})&=&\int \frac {d^3 {\bf q}}{(2\pi)^3} \frac{q_x^2}{[q_x^2+
 (\xi' q_{\perp}^2)^2]^2} (1-\cos({\bf q}\cdot{\bf r})) \\
 & =&\frac 1 {16\pi \xi'}
  \left\{ \ln \left[ 1 +(\Lambda_{\perp} r_{\perp})^2\right] +
  E_1\left(\frac{r_{\perp}^2}{4|x|\xi'} \right) \right. \nonumber\\ 
&& \left. +  e^{-r_{\perp}^2/(4|x|\xi')} \right\}\,, \nonumber
\end{eqnarray}
where $\xi'=\xi v_{\perp}/v_x$, $r_{\perp}^2=y^2+z^2$ and
  $\Lambda_{\perp}$ is a momentum cut-off. A study of the limits of
  this function for $r_\perp \to \infty$ and $|x|\to \infty$ shows
  that its asymptotic behavior is well described by:
\begin{eqnarray}\label{eq:expression-F}
  F({\bf r}) \sim \frac{v_x}{16\pi v_{\perp} \xi} \ln \left(\frac{r_\perp^2
  +4(v_{\perp}|x|\xi/v_x)}{\Lambda_{\perp}^{-2}}\right).  
\end{eqnarray}
\noindent Therefore, we have for $r\to \infty$:
\begin{eqnarray}
\label{eq:smectic-like}
 B^{FS}({r})  = \kappa   \ln \left(\frac{r_\perp^2
  +4(v_{\perp}|x|\xi/v_x)}{\Lambda_{\perp}^{-2}}\right).
\end{eqnarray}
\noindent where:
\begin{eqnarray}\label{eq:kappa-def}
  \kappa=\frac{D Q^2 v_x}{16\pi \xi \rho_0^2 v_{\perp}}.
\end{eqnarray}
The full asymptotic correlation function is given by the sum of the
forward scattering  contribution, Eq.~(\ref{eq:smectic-like}),
and the backward scattering contribution 
given in Eq.~(51) of Ref.~\onlinecite{rosso_cdw_long} for the
case of a short-range disorder and non-local elasticity:
\begin{equation}\label{eq:BS-log-log}
  B^{BS}(\mathbf{r})=log(log (\mathrm{max}(\Lambda|x|,(\Lambda r_\perp)^2))).
\end{equation}
Obviously, the contribution of the backward scattering terms is
subdominant and can be neglected. 

\subsection{Analogy with smectics-A}
\label{sec:smectics-analog}

We note that the result Eq.~(\ref{eq:expression-F}) can
be obtained in  the  entirely  different context of liquid crystals. If we consider a
smectic-A liquid crystal, its elastic free energy reads\cite{degennes_liquid_crystals,landau_elasticity,chandrasekhar_smectic,pieranski_book2}:
\begin{eqnarray}
  {\cal F}_{\text{el.}}=\int d^3{\bf r}  \left[\frac 1 2 B (\partial_z u)^2 + \frac 1 2 k_{11}
  (\Delta_\perp u)^2\right], 
\end{eqnarray}
\noindent where $u$ represents the displacement of the smectic
  layers, $B$ is the compressibility, and $k_{11}$ measures the
  bending energy of the smectic layers. If we now assume a random
  compression force given by:
  \begin{eqnarray}
    {\cal F}_{\text{dis.}} &=&\int d^3{\bf r} \eta({\bf r}) \partial_z
    u({\bf r}),\\  
  \overline{\eta({\bf r}) \eta({\bf r'})} &=& D \delta({\bf r}-{\bf
    r'}), 
  \end{eqnarray}
a straightforward calculation shows that the displacement correlation
function $\overline{(u({\bf r})-u(0))^2}$ is given by
Eq.~(\ref{eq:expression-F}).  
Smectics-A with disorder have been considered previously in Ref.~\onlinecite{radzihovsky99_smectique} albeit
  with a different type of disorder coupling to $\nabla_\perp u$. This
  yields a displacement correlation function superficially similar to $F(r)$
   with $q_\perp^2$ replacing $q_x^2$ in the numerator. The random
   compression force, which is not natural in the smectic-A context,
   is thus easily realized with charge density wave systems. 

\section{Experimental implications}
\label{sec:discuss} 
In the preceding sections, we have shown  that the forward scattering terms
generated by charged impurities 
lead to smectic-like order in a charge density wave material. 
A frequently used technique to characterize positional correlations
 in CDW systems is x-ray diffraction.\cite{cowley_x-ray_cdw} In the
 present section, we provide a calculation of the x-ray intensity
 resulting from such a smectic-like order, and we provide a
 quantitative estimate of the exponent $\kappa$. 
   
\subsection{x-ray intensity} 

 The intensity of the x-ray
spectrum is given by \cite{guiner_xray}
\begin{eqnarray}
    \label{eq:Sdef}
    I({q}) = \frac{1}{L^3} \sum_{i,j} e^{-i{q}({R}_i-{R}_j)} \left\langle
\overline{f_if_j e^{-iq(u_i-u_j)}} \right\rangle.
\end{eqnarray}  
 $u_i$ is the atom displacement from the
equilibrium position $R_i$, 
$f_i$ represents the total
amplitude scattered by the atom at the position $i$ and depends exclusively on
the atom type.  We consider the simple case of a disordered crystal,
made of one kind of atoms, characterized by the scattering factor
$\overline{f}-\Delta f/2$, 
and containing impurities of scattering
factor $\overline{f}+\Delta f/2$.  Since we are interested in the behavior of the scattering
intensity near a Bragg peak ($q\sim K$),  we can use the continuum
approximation.\cite{rosso_cdw_long} In the case of the CDW, the lattice modulation is given by:
\begin{equation}
  \label{eq:modulation}
  u(\mathbf{r})=\frac{u_0}{Q} \partial_x \left[ \cos (Q x
  +\phi(\mathbf{r}))\right],  
\end{equation}
It is well known that the presence of a CDW in the compound is
associated with  the appearance of two asymmetric satellites at positions
$q\sim K\pm Q$ around each
Bragg peak.\cite{cowley_x-ray_cdw} The intensity profiles of these
satellites give access to the structural properties of the CDW. For
this reason a lot of work has been done to compute and measure these
intensities.
\cite{ravy_x-ray_whiteline,brazovskii_x-ray_cdwT,rouziere_friedel_cdw,rosso_cdw_short,rosso_cdw_long}
By expanding Eq.~(\ref{eq:Sdef}) for low $q(u_i-u_j)$, one finds an
expression of the x-ray satellite intensity comprising a part
$I_{\text{d}}$, which is symmetric under inversion around the Bragg
  vector $K$ and a part  $I_{\text{a}}$  which is antisymmetric under
    the same transformation.\cite{rosso_cdw_long} 
The symmetric part is given by the following correlation function: 
\begin{equation}
  \label{eq:symmetric}
 I_{\text{d}}({\bf q}) =\overline{f}^2 q^2  \int d^3\mathbf{r}\,  e^{-i \delta
 {\bf q}\cdot {\bf r}}
 \langle \overline{u({\bf r}/2) u(-{\bf r}/2)} \rangle,   
\end{equation}
and the antisymmetric part by: 
\begin{equation}
  \label{eq:asymmetric}
  \frac{I_{\text a}({\bf q})}{{\bf a}\cdot({\bf b}\times {\bf c})} =  2 q {\Delta f}\, \mathrm{Im} \int d^3{\bf r}\, 
  e^{-i \delta {\bf q}\cdot {\bf r} } \langle\overline
{\rho_{\text{imp}}(\mathbf{r}/2) u(-\mathbf{r}/2) }  \rangle,
\end{equation}
\noindent where: $\delta q = (q-K)\sim Q$, and ${\bf a}\cdot({\bf
  b}\times {\bf c})$ is the volume of the unit cell of the crystal. 
After some manipulations, Eq.~(\ref{eq:symmetric}) can be rewritten as:
\begin{eqnarray}
  \label{eq:sym-simplified}
   I_{\text{d}}(K + Q + k)  &=&  u_0^2 \overline{f}^2 K^2 \int d^3{\bf
 r}\, e^{-i \mathbf{k\cdot r} }
 C_{\text{d}}(\mathbf{r}) \,, \\
I_{\text{a}}(K + Q + k)  &=& - \overline{f} K u_0 \Delta f \sqrt{{\cal
 N} D} \int  d^3{\bf
 r}\, e^{-i \mathbf{k\cdot r} } 
 C_{\text{a}}(\mathbf{r}) \nonumber 
\end{eqnarray}
\noindent where ${\cal N}$ is the number of impurities in the unit
 cell,  and: 
\begin{eqnarray}
   \label{eq:Cd}
 C_{\text{d}}(\mathbf{r})  &=& \left\langle \overline
 {e^{i(\phi(\mathbf{r}/2)-\phi(-\mathbf{r}/2))}} \right \rangle \,, \\
                  &=& C^{\text{F.S}}_{\text{d}}(\mathbf{r})
 C^{\text{B.S}}_{\text{d}}(\mathbf{r})\,,  \\ 
\label{eq:Ca} 
 C_{\text{a}}(\mathbf{r})  &=& \chi(\mathbf{r})  C_{\text{d}}(\mathbf{r}),
\end{eqnarray}
\noindent where $\chi(\mathbf{r})$ is defined by Eq. (33) of
Ref.~\onlinecite{rosso_cdw_long}. It is easy to show, using  this
 definition and the
 statistical tilt symmetry  that $\chi(r)$ is independent of the
 forward scattering disorder. In Eq.~(\ref{eq:Cd}), 
  $C^{\text{B.S}}_{\text{d}}$ is the backward scattering
 contribution which has been obtained in
 Ref.~\onlinecite{rosso_cdw_long}, and $C^{\text{F.S}}_{\text{d}}$ is the
 forward scattering contribution, given by: 

\begin{eqnarray}\label{eq:ccorr}
 C^{\text{F.S}}_{\text{d}}(\mathbf{r})=\left(\frac{\Lambda_{\perp}^{-2}}{r_{\perp}^2
 + 4 (v_{\perp}|x|\xi/v_x)}\right)^{\kappa},    
\end{eqnarray}
\noindent where we have used Eq.~(\ref{eq:smectic-like}), assuming a
 Gaussian disorder. Using Eq.~(\ref{eq:BS-log-log}), one sees that the
  term $C^{\text{B.S}}_{\text{d}}$ gives only 
 a logarithmic
 correction to (\ref{eq:Cd}). As a result, the symmetric structure
 factor $I_{\text{d}}$ is dominated by the contribution of the forward
 scattering disorder. 
To obtain the structure factor,  we  Fourier transform Eq.~(\ref{eq:ccorr}) to obtain
\begin{eqnarray}
  I_{\text{d}}({\bf q})&=&\int  d^2{\bf r_{\perp}} e^{i{\bf q_{\perp}
 r_{\perp}}} 
 \left(\frac{\Lambda_{\perp}^{-1}}{ r_{\perp}}\right)^{2\kappa}
  \frac{ r_{\perp}^2 v_x }{2\xi v_{\perp}} \nonumber
 \\ && \times \int_0^{+\infty}  \frac{du}{(1+u)^\gamma} \cos
  \left(\frac{|q_x|  r_{\perp}^2 v_x}{4\xi v_{\perp}} u \right)  \nonumber
\end{eqnarray}

\begin{figure}[htbp]
  \centering
  \includegraphics[width=\figwidth] {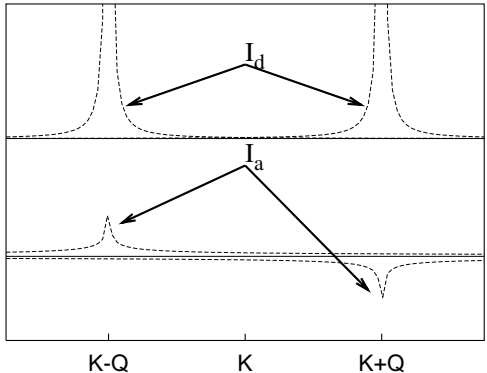} 
  \caption{A sketch of the x-ray intensity in a CDW with Coulomb
    elasticity and charged impurities for $K,Q$ parallel to the chain
    direction. We have taken $\kappa=0.5$ in the expressions of $I_a$
    and $I_d$.}
  \label{fig:peaks}
\end{figure}

Using the  following relation,
\begin{eqnarray}
 \int_0^{+\infty} \frac{du}{(1+u)^\gamma} \cos (\lambda u) = \frac
 {\lambda^{\gamma-1}} 
 {\Gamma(\gamma)} \int_0^{+\infty} dv
 e^{-v\lambda}\frac{v^\gamma}{v^2+1}    
\end{eqnarray}
we finally obtain
\begin{eqnarray}
  I_{\text{d}}({\bf q})&=&\frac{\pi (|q_x| \Lambda_{\perp}^{-1})^{\kappa-2}} {2^{2(\kappa -1)}
  \Gamma(\kappa)} \frac{  \Lambda_{\perp}^{-\kappa-2}}{ {(\xi
  v_{\perp}/v_x)^{\kappa-1}}} \nonumber \\ && \times
  \int_0^{+\infty} dw \frac{w^{1-\kappa}}{w^2 +1} e^{-w \frac{(\xi v_{\perp}/v_x)
  q_{\perp}^2}{2|q_x|}},  
\end{eqnarray}
\noindent so that $I_{\text{d}}({\bf q}) \sim (|q_x|)^{\kappa-2}$ for
$q_{\perp}^2 (\xi v_{\perp}/v_x) \ll |q_x|$ and $I_{\text{d}}({\bf q}) \sim
(|q_{\perp}|)^{2(\kappa-2)}$ otherwise.  The intensity $I_d({\bf q}=0)$ is divergent for
for $\kappa<2$ but is finite   for $\kappa>2$, i.e. for strong disorder.
 Next, we turn to the evaluation of
$I_{\text{a}}$. From Ref.~\onlinecite{rosso_cdw_long}, we know that
$\chi(r)\sim 1/x$ when $x\xi \gg r_\perp^2 $ and $\chi(r)\sim 1/r_\perp^2$ when
$|x|\xi\ll r_\perp^2$. This implies that $I_a$ is subdominant in comparison with
$I_d$. In particular, $I_{\text{a}}({\bf q}) \sim (|q_x|)^{\kappa-1}$ for
$q_{\perp}^2 (\xi v_{\perp}/v_x) \ll |q_x|$ and $I_{\text{a}}({\bf q}) \sim
(|q_{\perp}|)^{2(\kappa-1)}$ otherwise. We illustrate the behavior of
the x-ray intensities on Fig.~\ref{fig:peaks}. 

We note that these intensities are remarkably similar to
those of a disorder-free smectic-A liquid
crystal\cite{caille_smectic_xray} 
at  positive temperature.  In fact, the expression of the exponent
$\kappa$ Eq.~(\ref{eq:kappa-def}) is analogous to the expression
(5.3.12) in Ref.~\onlinecite{chandrasekhar_smectic}, with  the disorder
strength $D$ playing the role of the temperature $k_B T$ in the smectic-A liquid crystal.  
 
\subsection{Estimate of the exponent $\kappa$} 
 Let us turn to an  estimate of the
exponent $\kappa$ appearing in the intensities to determine whether
such smectic-like intensities are indeed observable in experiments. 
To do this, we  first need to determine whether Coulomb interactions are
unscreened by comparing the screening length with $\xi$ given by
Eq.~(\ref{eq:xi-definition}). This question is relevant only to a
material with a full gap, in which free uncondensed electrons cannot
screen charged impurities. A good candidate is  the blue bronze material
$\mathrm{K_{0.3}MoO_3}$ which has a full gap, and is well
characterized experimentally. 
We now evaluate the quantity $\xi$ for this material. 
Using the parameters  of Ref.~\onlinecite{pouget_bronzes}:
\begin{eqnarray}
  n_c=10^{20}\text{chains}/m^2,\\
  v_F=1.3\times 10^5 m.s^{-1}, \\ 
  \rho_0=3 \times 10^{27} e^-/m^3, \\ 
  Q=6\times 10^9 m^{-1},
\end{eqnarray}
and a relative permittivity of $\epsilon_{\mathrm{K_{0.3}MoO_3}}=1$,
so that $\epsilon$ in Eq.~(\ref{eq:xi-definition}) is equal to the
permittivity of the vacuum, we obtain: $\xi \simeq 5$\AA. Therefore,
the screening length can be large compared to $\xi$ at low temperature, and 
we expect that Coulombian effects will play an important role in this
material. We can use this value of $\xi$ to evaluate the exponent
$\kappa$ in Eq.~(\ref{eq:kappa-def}).  For the doped material $\mathrm{K_{0.3} Mo_{1-x} V_x O_3}$,
we find that the disorder strength can be expressed as a function of the doping and obtain:  
\begin{equation}\label{eq:banal}
  D=x(1-x)\frac{\#(\text{Mo atoms/unit cell})}{\mathbf{a}\cdot(\mathbf{b}\times\mathbf{c})}.
\end{equation}
This formula is derived in App.~\ref{app:disorder-strength}.      
For the crystal parameters, $a=$18.25\AA, $b=$7.56\AA, $c=$9.86\AA,
$\beta=$117.53$^o$ \cite{sato_xray_kmoo3}, with 20 Molybdenum
atoms per unit cell, and a doping  $x=3\%$,  the disorder strength $D=4.8\times 10^{26}
\mathrm{m}^{-3}$.  Moreover, using the experimental bounds of the velocities: $3.6\times 10^2 m.s^{-1}<v_\perp<1.6\times
10^3 m.s^{-1}$ and $v_x=3.7\times 10^3 m.s^{-1}$, we find that
$\kappa$ is in the range $[0.16-0.8]$. Therefore, the smectic-like
order should be observable in x-ray diffraction measurements on this material.

\section{Conclusion}
In this paper, we have introduced a decomposition of the 
 disorder induced by charged impurities  in terms  of the reciprocal
lattice vectors of a periodic charged elastic system. Using this decomposition,
we have shown that only the long wavelength (forward scattering)
component of the disorder was long-range correlated. Components with
 wavevectors commensurate with the reciprocal lattice of the periodic
 elastic system remain short ranged. The latter can thus be treated
 with the standard techniques developed for impurities producing
 short range forces.\cite{giamarchi_vortex_long} We find that  only the
 forward scattering is affected by the long-range character of the
 forces created by charged impurities. Due to the statistical tilt
 symmetry, this implies that only the statics of the periodic elastic
 system is modified by Coulombian disorder.    
This has allowed us  to obtain a full picture of the statics  
of charge density wave
systems in $d=3$  in the presence of charged and neutral impurities. The results
are summarized in table~\ref{tab:summary}.  
\begin{table}
\begin{center}
\begin{tabular}{ccc} 
\hline
Elasticity &\multicolumn{2}{c}{Disorder}
\\  & Short range  & Long range \\
\hline
Local  & $I_d(q)\sim q^{\eta-3}$ & $I_d(q)\sim q^{-3}$ \\
Non-local & Unphysical & \begin{tabular}{c} $I_d (q_x)\sim |q_x|^{\kappa-2}$ \\
$I_d(q_\perp)\sim |q_\perp|^{2(\kappa-2)}$ \\ \end{tabular} \\ \hline
\end{tabular}
\end{center}
\caption{The different cases with short range and long range random
  potential and elasticity. $\eta\simeq 1$ is the Bragg Glass
  exponent\cite{eta_exponent_variation}. $\kappa$ is defined in
  Eq.~(\ref{eq:kappa-def}).}
\label{tab:summary}  
\end{table} 
A remarkable result is that in the case of charged impurities in a
 system with unscreened Coulomb elasticity, the x-ray intensity turns
 out to be identical to that produced by thermal fluctuations in a
 smectic-A liquid crystal\cite{caille_smectic_xray}, with the disorder
 strength playing the role of an effective temperature. This behavior
 of the scattering intensity should be observable in the blue bronze
 material K$_{0.3}$MoO$_3$ doped with charged impurities such as
 Vanadium.  

 \begin{acknowledgments}
   This work was supported in part by the Swiss National Fund under
MANEP and division II. A. R. thanks the University of Geneva for
hospitality and support. We thank S. Brazovskii for enlightening
discussions. 
 \end{acknowledgments}

\appendix

\section{Decomposition of the Coulomb potential}
\label{app:decomp-coulomb}
In this appendix, we give more details on the decomposition of the Coulomb
potential using infinitely differentiable functions of compact
support.\cite{gelfand64_partition}  First,
let us discuss a simple decomposition in 1D.  
We consider the function $F_\epsilon(x)$ such that:
\begin{eqnarray}
  F_\epsilon(x)&=&F_\epsilon(-x)  \nonumber\\
  F_\epsilon(x)&=&1,\, 0 \le x \ge 1 \nonumber\\ 
  F_\epsilon(x)&=& \frac 1 2 \left\{ 1-\tanh\left[\frac{2
  (1-x)}{(x-1)^2+\epsilon^2}\right] \right\},\, |x-1|< \epsilon \nonumber \\
  F_\epsilon(x)&=&0,\,   x>1+\epsilon
\end{eqnarray}
It is easy to check that $F_\epsilon$ is continuous, infinitely
differentiable, and that:
\begin{eqnarray}
 \sum_{n=-\infty}^\infty F_\epsilon(x-2n)=1. 
\end{eqnarray}

Applying this formula to a one dimensional reciprocal space, we obtain:
\begin{eqnarray}
\label{eq:partition-unity}
  \sum_{n_x=-\infty}^\infty F_\epsilon\left(\frac a \pi \left(q-\frac
 {2\pi n}{a}\right)\right)=1,
\end{eqnarray}
\noindent i.e. we have constructed explicitly a partition of the
 unity.\cite{gelfand64_partition}  The generalization to a cubic lattice in
 a three dimensional space is obvious:
\begin{eqnarray}
&& \sum_{n_x,n_y,n_z= -\infty}^\infty F_\epsilon\left(\frac a \pi
 \left(q_x-\frac {2\pi n_x}{a}\right)\right) 
 F_\epsilon\left(\frac a \pi \left(q_y-\frac {2\pi
 n_y}{a}\right)\right)  \nonumber\\ && \times F_\epsilon\left(\frac a \pi \left(q_z-\frac {2\pi
 n_z}{a}\right)\right)= \sum_{\bf G} F^{3D}_\epsilon({\bf q}-{\bf G})
 =1. \nonumber \\ 
\end{eqnarray}
The function $F^{3D}_\epsilon$ has a compact support, and vanishes rapidly
outside the Brillouin zone.

\begin{figure}
\centerline{\includegraphics[width=8 cm]{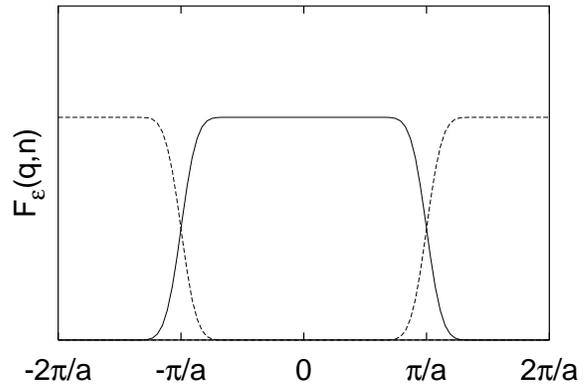}}
\caption{The graph of the function $F_\epsilon(qa/\pi)$, of compact support
  for a value of $\epsilon=\ldots$ (solid line). Dotted lines
  represent  $F_\epsilon(qa/\pi \pm 2)$. We can graphically check the
  validity of Eq.~(\ref{eq:partition-unity}). } 
\label{fig:compact}  
\end{figure}

\section{Contribution of the oscillating components of the density to
  the elastic Hamiltonian in the presence of Coulomb interaction} 
\label{app:backscatt}
 \begin{figure}[htbp]
    \centering
    \includegraphics[width=8cm]{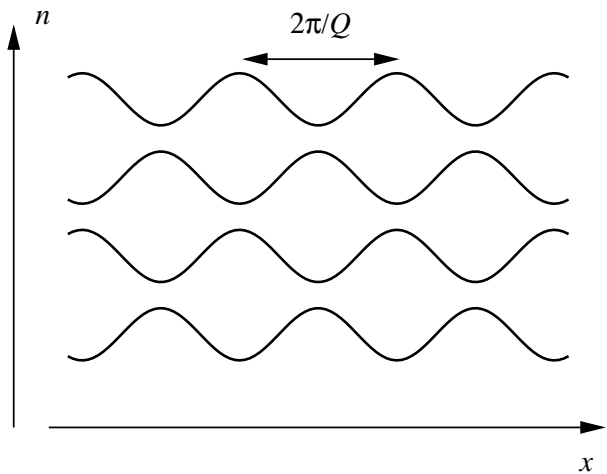} 
    \caption{The ground state of an array of CDWs coupled by a repulsive interaction. }
    \label{fig:coupled}
  \end{figure}

In this appendix, we calculate the contribution of the oscillating
terms to the Hamiltonian of the charge density wave with unscreened
Coulomb interactions, Eq.~(\ref{eq:electr}), and show that they
only induce  corrections to the  short range elasticity. Inserting the
expression of the density, Eq.~(\ref{eq:rhoexp}), in
Eq.~(\ref{eq:coulombenergy}), the contribution of the
oscillating component of wavevector $Q$ is given by,
\begin{eqnarray}
  H_{C}^{\text{osc.}}&=&\frac{e^2 \rho_1^2}{2 \epsilon}
  \sum_{\mathbf{n},\mathbf{n'}}  \int dx dx'
  V_{Q}(x-x',\mathbf{n-n'})\nonumber \\ &&\times \cos
  [\phi(x,\mathbf{n})-\phi(x',\mathbf{n'})],   
\end{eqnarray}
\noindent where we have reestablished the discrete character of the
  transverse dimension ${\bf y}$. 
Both the intrachain ($\mathbf{n}=\mathbf{n'}$) and the
  interchain contributions ($\mathbf{n}\ne\mathbf{ n'}$) are short
  ranged. Let us first
  consider the case of $\mathbf{n}\ne\mathbf{ n'}$.  
We have to compute integrals:
\begin{eqnarray}
V_{Q}(x,\mathbf{n})= \int \frac{d^{d}q}{(2\pi)^d} F_\epsilon^{3D}({\bf
   q}) \frac{e^{i (q_x x + \mathbf{q_\perp}
   \cdot \mathbf{n})}}{(Q+q_x)^2 + q_\perp^2},  
\end{eqnarray}
\noindent To evaluate the above integral in closed form, we need to
   make some approximations. Since $F_\epsilon$ vanishes for
   $Q+q_x=0$, we can neglect $q_x$ compared to $Q$ in this
integral.  Then,  we can extend the integration over the whole
   reciprocal space without encountering any singularity. The $q_x$
   integration produces a $\delta(x)$  function, and the
$q_\perp$ integration gives an exponential in 2D and a modified Bessel
function $K_0$ in 3D.
 In the two-dimensional case, we find that:
  \begin{eqnarray}
    V_{Q}(x,\mathbf{n})= \frac{e^{-Q\ell_y n}}{2Q} \delta(x),  
  \end{eqnarray}
\noindent  
and in the three-dimensional case:
\begin{eqnarray}\label{eq:shortrange-3d}
   V_{Q}(x,\mathbf{n})= \frac{K_0(Q\sqrt{(n_y \ell_y)^2 +(n_z
   \ell_z)^2})}{2\pi} \delta(x). 
\end{eqnarray}
\noindent
where, $\ell_y$ and $\ell_z$ are interchain spacings. Due to the exponential decay of the interchain interaction with
the distance, it is justified to neglect interchain interactions beyond
nearest neighbors. The logarithmic divergence in
Eq.~(\ref{eq:shortrange-3d}) for ${\mathbf{n}}=0$ 
is an artefact of the approximation we
make when we integrate over the entire reciprocal space instead of the
first Brillouin zone. A more refined estimate yields a finite, short
ranged intrachain contribution. The short range contribution in the
electrostatic energy thus reads: 
\begin{eqnarray}\label{eq:coulomb-final-sr}
  H^{\text{osc.}}_C&=&\sum_{\alpha=y,z} J_\alpha \int dx \sum_{\mathbf{n} } \cos
  [\phi(x,\mathbf{n}+\mathbf{e}_\alpha)-\phi(x,\mathbf{n})] +\nonumber
  \\ && +\sum_{\mathbf{n} } \int V_Q(x-x',0) dx dx' \cos
  [\phi(x,\mathbf{n})-\phi(x,\mathbf{n})]\nonumber, \\
\end{eqnarray}
\noindent where: 
\begin{equation}\label{eq:def-J}
  J_\alpha = \frac{e^2 \rho_1^2}{2\pi\epsilon} K_0(Q \ell_\alpha).  
\end{equation}
 In Eq.~(\ref{eq:coulomb-final-sr}), we expand $ \cos (
  \phi(x,\mathbf{n})-\phi (x',\mathbf{n})) \sim
  1-(x-x')^2/2 (\partial_x \phi(x,\mathbf{n}))^2$ 
to show that the backscattering
  term reduces to short ranged elastic forces.
 Making
  $\phi(x,\mathbf{n})={\bar \phi}(x,\mathbf{n}) +n_y\pi+n_z \pi$, we
  can make the sign in front of $J_\alpha$ negative, and obtain the
  ground state for ${\bar \phi}=0$. 
 In its ground state, the CDW
  is out of phase on two nearest neighbor chains. This ground state is
  represented on Fig.~\ref{fig:coupled}. The excitations above this
  ground state are described by the Lagrangian~(\ref{eq:3D-CDW})
  derived in App.~\ref{app:lagrangian-cdw}.

\section{Zero temperature limit of the charge density in a CDW}
\label{app:cdw-density}
In the present appendix, we discuss the zero temperature limit of the
expression of the charge density in a CDW. 
Let us consider
the form  of the charge density in the presence of  a non-uniform
$\phi$. It is
given\cite{fukuyama_cdw_pinning,lee_coulomb_cdw,lee_depinning,rice_lee_cross}
by 
the expression:
\begin{eqnarray}
  \label{eq:charge-phi}
  \rho({\bf r}) = \rho_0+ \frac {\rho_0\bar{\rho}_c} {Q} \partial_x \phi({\bf r}) +\rho_1
  |\psi| \cos (Qx+\phi({\bf r}))   
\end{eqnarray}
\noindent where $\rho_0$ is the average electron density, $\rho_1$ is
  the condensate amplitude at $T=0$, $|\psi|$ takes into account the
  reduction of CDW order by thermal fluctuations ($|\psi|=1$ at
  $T=0$), the factor $\bar{\rho}_c$ takes into account the presence of
  non-condensed electrons\cite{rice_lee_cross} at finite temperature (at $T=0$,
  $\bar{\rho}_c=1$)  and
  $Q=2k_F$. Using the relation $k_F=\frac \pi 2 \rho_0$, valid in
  a one-dimensional system, one can
  see that this relation simplifies (at $T=0$)  to:
  \begin{eqnarray}
\label{eq:charge-phi-simpler} 
    \delta \rho = \frac{\partial_x \phi}{\pi}.
  \end{eqnarray}
In the paper, we consider temperatures very low compared
to the Peierls transition temperature, and we take $|\psi|=1$,
$\bar{\rho}_c=1$. This yields Eq.~(\ref{eq:rhoexp}).

\section{Derivation of the Hamiltonian of  a three-dimensional charge density wave}\label{app:lagrangian-cdw}
In this appendix, we provide a derivation of the Hamiltonian of a
three dimensional CDW starting from the original
Fukuyama one dimensional description.
The Lagrangian of a CDW in a single chain is given
by:\cite{fukuyama_cdw_pinning} 
\begin{eqnarray}
  \label{eq:CDW-lagrangian}
  {\cal L}=\frac {\hbar v_F} {4\pi} \int dx \left[ \frac{(\partial_t
  \phi)^2}{v_\phi^2} -(\partial_x \phi)^2\right]  
\end{eqnarray}
Obviously, this Lagrangian describes phason waves propagating with the
velocity $v_\phi$.  $v_F$ is the Fermi velocity of the
electrons forming the CDW and  $\phi$ is the phase of the CDW. 
We define an effective mass $m^*$ by:
\begin{eqnarray}
  \label{eq:mstar-def}
  \frac {v_F^2}{v_\phi^2} = \frac{m^*}{m_e}  
\end{eqnarray}
where $m_e$ is the electron mass.

In a three-dimensional CDW with screened Coulomb interactions, 
the chains are coupled by a backscattering interaction.   
The resulting Lagrangian reads:
\begin{eqnarray}
  \label{eq:3D-CDW}
 {\cal L}&=&\frac {\hbar v_F} {4\pi} \sum_{\mathbf{n}} \int dx \left[ \frac{(\partial_t
  \phi)^2}{v_\phi^2} -(\partial_x \phi)^2 \right](x,\mathbf{n}) 
  \\ &&+ \frac 1 2
  \sum_{\langle \mathbf{n},\mathbf{n'} \rangle}
  J(\mathbf{n},\mathbf{n'}) \int dx \cos [\phi(x,\mathbf{n})
  -\phi(x,\mathbf{n'})], 
  \nonumber      
\end{eqnarray}
\noindent where $J(\mathbf{n},\mathbf{n'})$ is short-ranged and is
given   by Eq.~(\ref{eq:def-J}).   
Expanding the cosines,
\begin{eqnarray}
  &&\cos[\phi(x,\mathbf{n}) -\phi(x,\mathbf{n+\mathbf{e}_y})] =
  \nonumber \\ && =1 -
  \frac{(\phi(x,\mathbf{n}) -\phi(x,\mathbf{n}+\mathbf{e_y}))} 2 + o(\nabla_y^2)
  \nonumber \\ 
   &&= 1 - \frac{\ell_y^2}{2} (\partial_y \phi)^2 + o(\partial_y \phi)^2,   
\end{eqnarray}
and defining,
\begin{eqnarray}
  \frac{\hbar v_F}{4\pi} \frac{v_{y,z}^2}{v_\phi^2} &=& J_{y,z} \frac{\ell_{y,z}^2}{2}, 
\end{eqnarray}
the Lagrangian  in Eq. (\ref{eq:3D-CDW}) can be rewritten as\cite{lee_coulomb_cdw}:
\begin{eqnarray}
  {\cal L} &=& \sum_{\mathbf{n}} \frac{\hbar v_F}{4\pi} \int dx \left[ 
\frac {(\partial_t \phi)^2}{v_\phi^2} -(\partial_x \phi)^2  
\right. \nonumber \\ && \left . -
\frac{v_y^2}{v_\phi^2} (\partial_y  \phi)^2 - 
\frac{v_y^2}{v_\phi^2} (\partial_y  \phi)^2   \right] 
\end{eqnarray}
\noindent
The
sum over lattice sites in the transverse direction can be replaced by
an integral, by writing:
\begin{eqnarray}
   {\cal L} &=& \frac{\hbar v_F}{4\pi \ell_y \ell_z } \int
   d^3{\mathbf{r}}  \left[ 
\frac {(\partial_t \phi)^2}{v_\phi^2} -(\partial_x \phi)^2  \right.
   \nonumber \\ && \left. -
\frac{v_y^2}{v_\phi^2} (\partial_y  \phi)^2 - 
\frac{v_y^2}{v_\phi^2} (\partial_y  \phi)^2   \right]  
\end{eqnarray}The phason dispersion is now $\omega(q)^2 = v_\phi^2 q_x^2 + v_y^2 q_y^2 + v_z q_z^2$.
The momentum conjugate to $\phi$ is obtained by the usual relation:
\begin{equation}
  \label{eq:def-pi}
  \Pi=\frac{\delta {\cal L}}{\delta \left( \partial_t \phi\right)}
  = \frac{\hbar v_F}{2\pi v_\phi^2 \ell_y \ell_z} \partial_t \phi, 
\end{equation}
\noindent yielding the Hamiltonian:
\begin{eqnarray}
  \label{eq:full-quantum}
  H&=& \frac{\hbar v_F}{4\pi \ell_y \ell_z } \int d^3 \mathbf{r}
  \left[\frac{4\pi^2 v_\phi^2 (\ell_y \ell_z)^2}{\hbar^2  v_F^2} \Pi^2
    + (\partial_x \phi)^2  \right. \nonumber \\ && \left.  +
\frac{v_y^2}{v_\phi^2} (\partial_y  \phi)^2 +
\frac{v_y^2}{v_\phi^2} (\partial_y  \phi)^2 \right]. 
\end{eqnarray}
\noindent From the Hamiltonian Eq.~(\ref{eq:full-quantum}), it is
  straightforward to obtain the Debye-Waller factor associated with the
  zero point fluctuations of the phase $\phi$. In the isotropic case,
  $v_y=v_z=v_\phi$, one finds that $\langle \cos \phi(x) \rangle_{T=0}
  \sim e^{-C(m_e/m^*)^{1/2}}$, where $C\sim \pi/4$ is a dimensionless constant
  of order $1$.  
Due to the smallness of the ratio $m_e/m^*\sim 10^{-2}$, the zero
  point motion can be neglected, and the kinetic term $\propto \Pi^2$ in
  Eq.~(\ref{eq:full-quantum}) can be dropped. This leads to the
  Hamiltonian~(\ref{eq:hamiltonian_cdw_sr}).

\section{Estimation of the disorder strength}
\label{app:disorder-strength}

Here, we give an estimation of the disorder strength $D$
in doped $\mathrm{KMo_{1-x}V_xO_3}$. We assume  a binomial distribution of Vanadium impurities on the Molybdenum
sites. The Vanadium impurities carry an extra electron compared to the
Molybdenum ions. The resulting charge density fluctuation reads:
\begin{equation}
  \label{eq:delta-rho}
  \delta\rho(\mathbf{r})=\sum_{i,\alpha} (x-\sigma_{i,\alpha})
  \delta(\mathbf{r}-\mathbf{R}_{i,\alpha}), 
\end{equation}
where $i$ is the index of the cell and $\alpha$ is the index of the
Molybdenum site in a given cell.
$\sigma_{i,\alpha}=0$ if the site is occupied by a Molybdenum
ion, and  $\sigma_{i,\alpha}=1$ if it is occupied by a Vanadium impurity.
By construction, the expectation
value of $\delta\rho(\mathbf{r})$ is zero. We estimate the second
moment of $\delta\rho(\mathbf{r})$ as:
\begin{eqnarray}
&&  \overline{\delta\rho(\mathbf{r})\delta\rho(\mathbf{r'})} = \nonumber \\
&& = \sum_{i,j,\alpha,\beta} \overline{(x-\sigma_{i,\alpha})
  (x-\sigma_{j,\beta})} \delta(\mathbf{r}-\mathbf{R}_{i,\alpha})
  \delta(\mathbf{r'}-\mathbf{R}_{j,\beta})\nonumber \\ 
&&  = x(1-x)  \sum_{i,\alpha} \delta(\mathbf{r}-\mathbf{R}_{i,\alpha})
  \delta(\mathbf{r'}-\mathbf{r}),    
\end{eqnarray}
\noindent where we have used the property that $\overline{(x-\sigma_{i,\alpha})
  (x-\sigma_{j,\beta})} =\delta_{i,j}\delta_{\alpha,\beta}
  \overline{(x-\sigma_{j,\beta})^2}$. The expectation value of 
  $   \sum_{i,\alpha} \delta(\mathbf{r}-\mathbf{R}_{i,\alpha})$ 
   is simply the number of Molybdenum ions per unit volume,
leading to the formula Eq.~(\ref{eq:banal}). 

\begin{thebibliography}{10}
\bibitem{eta_exponent_variation}
The exponent $\eta$ is universal and the best prediction is given by FRG
  calculation $\eta \sim 1.1$. However also in the case where the elastic
  object has many components (i.e. for the vortex lattice) the possible
  variation of this exponent is still less than a percent. [S. Bogner, T. Emig
  and T. Nattermann Phys. Rev. B {\bf 63} 174501 (2001)].
\end{thebibliography}

\end{document}